\documentclass[conference]{IEEEtran}
\IEEEoverridecommandlockouts
\usepackage{verbatim}
\usepackage{longtable}
\usepackage{adjustbox}
\usepackage{listings}
\usepackage{xcolor}

\lstdefinelanguage{Solidity}{
  morekeywords={pragma, solidity, contract, struct, event, function, mapping, address, bytes, bytes32, bool, external, public, require, emit, calldata, payable, return, returns, if, else, memory, keccak256},
  sensitive=true,
  morecomment=[l]{//},
  morecomment=[s]{/*}{*/},
  morestring=[b]"
}

\usepackage{amsmath}
\usepackage{amsfonts}
\usepackage{amssymb}
\usepackage{multirow}
\usepackage{amsmath,amssymb,amsfonts}
\usepackage{fancyhdr}

\usepackage{graphicx}
\usepackage{textcomp}
\usepackage{xcolor}
\usepackage{enumitem}
\usepackage{float}
\usepackage{graphicx}
\usepackage[normalem]{ulem}
\usepackage{url}
\useunder{\uline}{\ul}{}
\usepackage{cite}
\usepackage{amsmath,amssymb,amsfonts}
\usepackage{graphicx}
\usepackage{textcomp}
\usepackage{comment}
\usepackage{subcaption}

\usepackage{xcolor}
\def\BibTeX{{\rm B\kern-.05em{\sc i\kern-.025em b}\kern-.08em
    T\kern-.1667em\lower.7ex\hbox{E}\kern-.125emX}}
\usepackage{algorithm}
\usepackage[noend]{algpseudocode}

\begin{document}

\title{Rollback-Free Cross-Chain Atomicity Through Forward-Only Correction}

\author{\IEEEauthorblockN{Tahrim Hossain, Faisal Haque Bappy, Tarannum Shaila Zaman, Tariqul Islam}
\IEEEauthorblockA{
University of Maryland Baltimore County\\
Email: \{m482, fbappy1, zamant, mtislam\}@umbc.edu} 
}
\maketitle

\begin{abstract}

Blockchain platforms have grown into an ecosystem of independent networks, and a growing class of applications now requires smart contracts on separate chains to act as one. Such operations must be atomic, yet immutability makes this fundamentally harder: a confirmed transaction cannot be reversed, so the rollback on which classical atomic commitment protocols depend is unavailable. Two challenges follow. Contract state must be held across an operation whose outcome is not yet known, and each chain's execution outcome must be established even though no chain can observe another. In response, we introduce a framework that achieves atomicity through forward-only correction, resolving incomplete operations with new on-chain transactions rather than reversal. The framework bounds how long contract state is held and confines contention to the state an operation touches, and it establishes outcomes from an on-chain record of what each chain executed, without relying on any single coordinating party. This work lays the foundation for atomic coordination of general smart contract operations across heterogeneous blockchains.

\end{abstract}


\section{Introduction}
A smart contract is a self-executing program that encodes an agreement and enforces its terms automatically on a blockchain, without a trusted intermediary. Once the contract executes, its state changes are recorded on the blockchain permanently and cannot be reversed \cite{szabo1996smart, buterin2013ethereum}. These contracts now support a range of production applications, including financial settlement \cite{sharma2024past}, supply chain tracking \cite{rahman2021framework}, healthcare data management \cite{ullah2025toward}, and decentralized governance \cite{fabrega2025voting}. While these applications were once confined to a single network, the ecosystem has since diversified into numerous blockchain platforms, each adopting distinct tradeoffs among throughput, cost, and governance \cite{nakamoto2008bitcoin}. These platforms remain isolated by design. Because each network establishes trust solely through its own consensus over its own state, a blockchain cannot natively verify the state of another, rendering cross-chain state inaccessible without additional trust assumptions.

This isolation is increasingly at odds with how applications operate. A growing class of operations spans several chains at once, such as a loan collateralized on one chain and issued on another, and binding their separate steps into one outcome requires atomicity: every chain commits its part, or none does. This is the atomic commitment problem, long studied in distributed databases \cite{ram1999distributed, zakhary2020atomic}. Classical protocols such as two-phase commit solve it through rollback, preparing changes tentatively and undoing them if coordination fails \cite{gray1992transaction, ozsu1999principles}. Blockchains break this assumption, because a confirmed transaction is permanent and cannot be undone. Removing rollback also sharpens two problems classical solutions handled through mechanisms blockchains lack: locks now guard economically valuable positions, so their duration carries direct cost, and a failure on one chain produces no signal the others can verify without additional trust.
We argue that atomicity under immutability requires abandoning reversal. If a committed effect cannot be undone, coordination must correct outcomes by moving forward, not backward. We develop this as rollback-free corrective action: rather than reverting a committed effect, a participant issues a new committed effect that compensates for it. Atomicity thus becomes a property of forward-only state transitions, preserving immutability while the operation still resolves to a single agreed result.

This paper makes three contributions.
\begin{itemize}
    \item We identify immutability as the fundamental barrier to atomic cross-chain smart contract execution.
    \item We formulate contention over contract state and the verification of cross-chain execution outcomes as the core challenges of coordination without rollback.
    \item We propose a framework that enforces atomicity across chains by journaling state changes and correcting incomplete operations with inverse state changes.
\end{itemize}

\section{System Architecture}

\begin{figure}[htbp!]
  \centering
  \includegraphics[width=0.9\columnwidth]{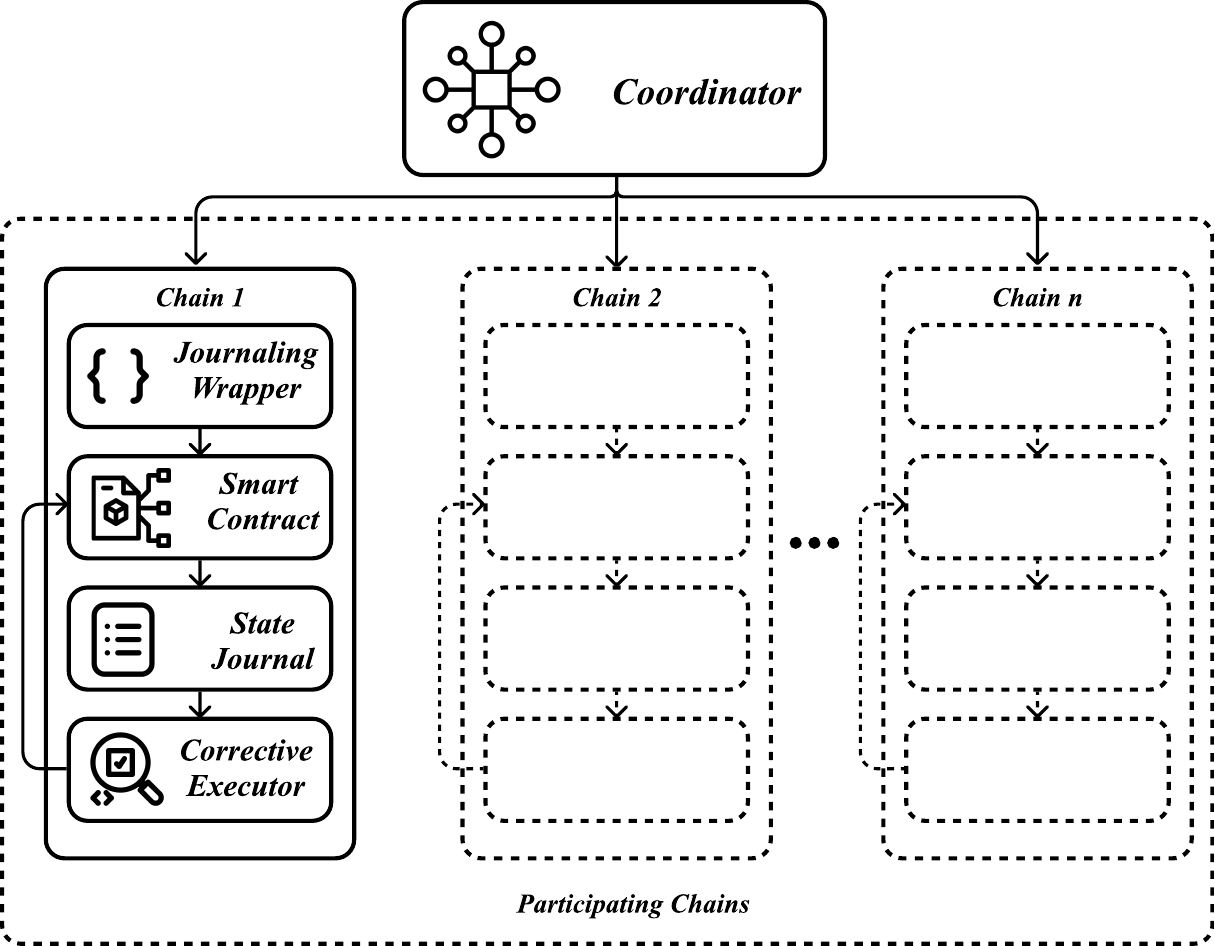}
  \caption{System architecture. One participating chain is shown in full; the remaining chains hold the same components.}
  \label{fig:archi}
\end{figure}

The framework consists of four components that carry a cross-chain operation from invocation to resolution, as illustrated in Fig.~\ref{fig:archi}. Together they initiate the operation, execute each participating contract's function call while recording the state it changes, store those records on-chain, and apply inverse state changes when the operation cannot complete. We describe these components and their interaction below.

\subsection{Coordinator}
The Coordinator is a distributed service replicated across independent nodes, none of which can act alone. It initiates a cross-chain operation by registering it with a unique identifier and the set of participating chains, then invoking the designated function on each. Each node independently observes the evidence a chain emits when it performs its part, and the nodes agree on the outcome through a quorum. The operation is finalized only if a quorum observes every chain reporting within the deadline, and correction is directed otherwise. Every directive carries a quorum certificate, which the contracts it invokes verify before acting. Distributing this role lets chains coordinate without observing one another directly, and without a single party deciding the outcome.

\subsection{Journaling Wrapper}
The Journaling Wrapper is deployed alongside each participating contract and mediates the function calls belonging to a cross-chain operation. On invocation, it reads the storage slots the target function will modify, executes the call, and reads those slots again once the call returns. The resulting pre-image and post-image pair, together with the operation identifier, forms a journal entry. The wrapper locks the affected slots for the duration of the operation, rejecting any other tracked call that would modify them until the operation resolves. Locking at slot granularity confines contention to the state an operation actually touches, leaving the rest of the contract available. The wrapper records the entry in the chain's on-chain journal and emits a \textit{StateRecorded} event, which serves as the on-chain evidence that this chain performed its part. The call itself executes and is confirmed like any other transaction, so the wrapper adds recording and locking without altering execution semantics.

\subsection{State Journal}
The State Journal is the on-chain store in which each chain records its journal entries, keyed by operation identifier. Each entry records the affected storage slots, their pre-images and post-images, the lock held over those slots and its expiry, and a status flag marking the entry as pending, finalized, or corrected. Entries are append-only. A correction appends a record rather than overwriting the original, so the journal preserves the full history of an operation and the chain retains both the original call and its correction. Retaining the pre-image on-chain is what makes correction possible without reversal, since the state preceding a call remains recoverable as data even though the transaction that changed it cannot be undone.

\subsection{Corrective Executor}
The Corrective Executor applies the inverse of a recorded state change when an operation cannot complete. On a correction directive from the Coordinator, it reads the pending journal entry for that operation, writes the recorded pre-image values back into the contract's state, releases the locks the entry holds, and marks it corrected. This restores the state that preceded the operation as a new confirmed transaction rather than as a reversal. Locks are also released on finalization, and any lock that reaches its expiry without a directive triggers correction on its own chain, so no operation holds state indefinitely. The executor accepts directives only from the registered Coordinator and only for entries in the pending state, which bounds the Coordinator's authority. It may correct an operation that did not complete, but it cannot alter state that no recorded call produced.

\section{Conclusion}
In this paper, we introduced a framework for atomic cross-chain smart contract execution that achieves atomicity without rollback. Rather than reversing committed effects, our approach records the state each function call changes and corrects incomplete operations by applying inverse state changes as new on-chain transactions, restoring consistency without violating immutability. By operating on general function calls rather than asset transfers alone, the framework extends atomic coordination to a broad class of cross-chain applications. In the future, we will focus on a full framework specification, formal analysis of its safety and liveness guarantees, and a prototype implementation to evaluate its overhead across heterogeneous chains.

\bibliographystyle{IEEEtran}
\bibliography{IEEEabrv,references}

\end{document}